\documentclass[twocolumn,aps,floats,preprintnumbers]{revtex4}
\usepackage{epsfig,amsmath,amsfonts,amssymb,euscript}
\begin{document}

\preprint{FERMILAB-PUB-08-310-T}
\preprint{MZ-TH/08-25}

\title{Origin of the Large Perturbative Corrections to Higgs Production at Hadron Colliders}

\author{Valentin Ahrens\,$^a$, Thomas Becher\,$^b$, Matthias Neubert\,$^a$, and Li Lin Yang\,$^a$
\vspace{0.2cm}}

\affiliation{$^a$\,Institut f\"ur Physik (THEP)\\
Johannes Gutenberg-Universit\"at Mainz\\ 
D-55099 Mainz, Germany\vspace{2mm}\\
$^b$\,Fermi National Accelerator Laboratory\\
P.O. Box 500, Batavia, IL 60510, U.S.A}

\begin{abstract}
The very large $K$-factor for Higgs-boson production at hadron colliders is shown to result from enhanced perturbative corrections of the form $(C_A\pi\alpha_s)^n$, which arise in the analytic continuation of the gluon form factor to time-like momentum transfer. These terms are resummed to all orders in perturbation theory using the renormalization group. After the resummation, the $K$-factor for the production of a light Higgs boson at the LHC is reduced to a value close to 1.3.
\end{abstract}

\maketitle

\section{Introduction}

The discovery of the Higgs boson is the most important goal of modern particle physics. The inclusive production cross sections for $pp\to H+X$ and $p\bar p\to H+X$ have been calculated a long time ago at next-to-leading order (NLO) in perturbation theory \cite{Dawson:1990zj,Djouadi:1991tka}, and since a few years results at next-to-next-to-leading order (NNLO) are available \cite{Harlander:2002wh,Anastasiou:2002yz,Ravindran:2003um}. Inclusive Higgs-boson production is thus one of the best studied processes from a theoretical perspective.

In view of this fact, it is uncomfortable that the behavior of QCD perturbation theory appears to be rather poor in this case. The $K$-factor for Higgs-boson production, defined as the prediction for the cross section normalized to the Born approximation, takes surprisingly large values. For the production of a light Higgs boson ($m_H<150$\,GeV) at the LHC, one typically finds $K\approx 1.7$--1.9 at NLO and $K\approx 2.0$--2.2 at NNLO. Also, the residual dependence on the renormalization and factorization scales remains significant even at NNLO. The standard argument that the large $K$-factor results from the accessibility of new production channels beyond the leading order, such as $q g\to Hq$ and $q\bar q\to H$, does not apply in this case, as these contributions to the cross section are known to be below 10\%. Also, the $K$-factor is not much reduced by soft-gluon resummation near the partonic threshold \cite{Catani:2003zt}.

In this Letter we show that the bulk of the large perturbative corrections to Higgs-boson production via gluon-gluon fusion originate from terms of the form $(C_A\pi\alpha_s)^n$ arising from the analytic continuation of the gluon form factor to time-like momentum transfer, and that these terms exponentiate to leading order.

\section{Time-Like gluon form factor}

The Higgs-boson production cross section at hadron colliders such as the Tevatron or LHC is dominated by the gluon-gluon fusion process $gg\to H$ via a top-quark loop. For a not too heavy Higgs boson, this process is well approximated by the effective local interaction \cite{Inami:1982xt}
\begin{equation}\label{Heff}
   {\cal L}_{\rm eff} = C_t(m_t^2,\mu^2)\,\frac{H}{v}\,
   G_{\mu\nu,a}\,G_a^{\mu\nu} \,,
\end{equation} 
where $v\approx 246$\,GeV is the Higgs vacuum expectation value, and the short-distance coefficient $C_t=\alpha_s/(12\pi)+{\cal O}(\alpha_s^2)$ is known to NNLO  \cite{Chetyrkin:1997iv} and has a well behaved perturbative expansion for $\mu\sim m_H$. The production cross section is related to the discontinuity of the product of two such effective vertices. It can be written as the convolution of a hard-scattering kernel with parton distribution functions. 

The large corrections we identify are due to virtual corrections to the effective $ggH$ interaction (\ref{Heff}) and arise from quantum corrections characterized by the scale $\mu\sim m_H$. These effects are described by a universal factor and affect differential distributions in same way as the total cross section. They can be factorized into a hard function $H(m_H^2,\mu^2)$, which is the square of the on-shell gluon form factor evaluated at time-like momentum transfer $q^2=m_H^2$, and with infrared divergences subtracted using the $\overline{\rm MS}$ scheme \cite{Becher:2006nr,Idilbi:2006dg,Becher:2007ty}. On a  technical level, the hard function appears as a Wilson coefficient in the matching of the two-gluon operator in (\ref{Heff}) onto an operator in soft-collinear effective theory (SCET) \cite{Bauer:2001yt,Bauer:2002nz}, in which all hard modes have been integrated out. This matching takes the form
\begin{equation}\label{CSdef}
   G_{\mu\nu,a}\,G_a^{\mu\nu}
   \to C_S(Q^2,\mu^2)\,Q^2\,g_{\mu\nu}\,
   {\EuScript A}_{n\perp}^{\mu,a}\,
   {\EuScript A}_{\bar n\perp}^{\nu,a} \,,
\end{equation}
where $Q^2=-q^2$ is (minus) the square of the total momentum carried by the operator. The fields ${\EuScript A}_{n\perp}^{\mu,a}$ and ${\EuScript A}_{\bar n\perp}^{\nu,a}$ are effective, gauge-invariant gluon fields in SCET \cite{Hill:2002vw}. They describe  gluons propagating along the two light-like directions $n,\bar n$ defined by the colliding hadrons. 

The two-loop expression for the Wilson coefficient $C_S$ can be extracted from the results of \cite{Harlander:2000mg}. We write
\begin{equation}\label{CSexp}
   C_S(Q^2,\mu^2) = 1 + \sum_{n=1}^\infty\,c_n(L)
   \left( \frac{\alpha_s(\mu^2)}{4\pi} \right)^n \! ,
\end{equation}
where $L=\ln(Q^2/\mu^2)$. The one-loop coefficient reads
\begin{equation}\label{c1c2}
   c_1(L) = C_A \left( -L^2 + \frac{\pi^2}{6} \right) ,
\end{equation}
and the result for the two-loop coefficient can be found in \cite{Idilbi:2006dg,inprep}. The hard function is given by the absolute square of the Wilson coefficient at time-like momentum transfer, 
\begin{equation}\label{Hdef}
   H(m_H^2,\mu^2) = \left| C_S(-m_H^2-i\epsilon,\mu^2) \right|^2 .
\end{equation}
The Wilson coefficient obeys an evolution equation, which reflects the renormalization properties of the effective two-gluon operator in SCET. It reads \cite{Becher:2006nr}
\begin{equation}\label{CSevol}
   \frac{dC_S(Q^2,\mu^2)}{d\ln\mu}
   = \left[ \Gamma_{\rm cusp}^A(\alpha_s)\,\ln\frac{Q^2}{\mu^2}
   + \gamma^S(\alpha_s) \right] C_S(Q^2,\mu^2) \,,
\end{equation}
where $\Gamma_{\rm cusp}^A$ is the cusp anomalous dimension of Wilson lines with light-like segments in the adjoint representation of $SU(N_c)$. It controls the leading Sudakov double logarithms contained in $C_S$ and is known to three-loop order \cite{Vogt:2004mw}. The single-logarithmic evolution is controlled by the anomalous dimension $\gamma^S$, which can be extracted from the infrared divergences of the on-shell form factor \cite{Becher:2006nr}. Using results from \cite{Moch:2005tm} it can be derived to three-loop order \cite{inprep}. The evolution equation (\ref{CSevol}) links the coefficients of the logarithmic terms in (\ref{CSexp}) to coefficients in the perturbative expansions of the anomalous dimensions and the QCD $\beta$-function. At one-loop order we have
\begin{equation}\label{c1general}
   c_1(L) = - \frac{\Gamma_0^A}{4}\,L^2
   - \frac{\gamma_0^S}{2}\,L + C_A\,\frac{\pi^2}{6} \,,
\end{equation}
where $\Gamma_0^A=4C_A$ and $\gamma_0^S=0$.

The Wilson coefficient at space-like momentum transfer has a well behaved expansion in powers of the coupling constant, if the renormalization scale is taken to be of order the natural scale, $\mu^2\sim Q^2$. For instance, with $N_c=3$ colors and $n_f=5$ light quark flavors, we find
\begin{equation}\label{CSeucl}
   C_S(Q^2,Q^2) = 1 + 0.393\,\alpha_s(Q^2) 
   - 0.152\,\alpha_s^2(Q^2) + \dots \,.
\end{equation} 
The nature of the expansion changes drastically when the same coefficient is evaluated at time-like momentum transfer $Q^2=-q^2-i\epsilon$. We then obtain
\begin{eqnarray}\label{CSbad}
   C_S(-q^2,q^2) 
   &=& 1 + 2.75\,\alpha_s(q^2) + (4.84+2.07i)\,\alpha_s^2(q^2) 
    \nonumber\\
   &&\mbox{}+ \dots \,.
\end{eqnarray} 
The expansion coefficients are more than an order of magnitude larger than in the space-like region. The origin of this effect is that the Sudakov (double) logarithms contained in the coefficients $c_n(L)$ in (\ref{CSexp}) give rise to $\pi^2$ terms when we analytically continue $L\to\ln(q^2/\mu^2)-i\pi$. For the hard function entering the Higgs-boson production cross section, this implies
\begin{eqnarray}\label{Hres}
   H(m_H^2,m_H^2) 
   &=& 1 + 5.50\alpha_s(m_H^2) + 17.24\alpha_s^2(m_H^2) + \dots 
    \nonumber\\
   &=& 1 + 0.623 + 0.221 + \dots \,,
\end{eqnarray}
where the numerical estimates in the last line refer to the NLO and NNLO corrections for a Higgs-boson mass of 120\,GeV, and we use $\alpha_s(m_Z^2)=0.118$ as our normalization of the running coupling constant. These hard matching corrections account for the bulk of the $K$-factors found at NLO and NNLO.

The large expansion coefficients in the perturbative series for the Wilson coefficient in the time-like region can be avoided if we evaluate this coefficient at a {\em time-like\/} renormalization point, in which case (here and below, negative arguments of the running coupling are always understood with a $-i\epsilon$ prescription)
\begin{equation}\label{Ctimelike}
   C_S(-q^2,-\mu^2) = 1 + \sum_{n=1}^\infty\,c_n(L)
   \left( \frac{\alpha_s(-\mu^2)}{4\pi} \right)^n
\end{equation}
with $L=\ln(q^2/\mu^2)$ and the {\em same\/} expansion coefficients as in (\ref{CSexp}). We then obtain
\begin{equation}\label{CSgood}
   C_S(-q^2,-q^2) = 1 + 0.393\,\alpha_s(-q^2) 
   - 0.152\,\alpha_s^2(-q^2) + \dots
\end{equation} 
instead of (\ref{CSbad}). The perturbative series analogous to that in (\ref{Hres}) reads
\begin{equation}\label{Cgood}
   |C_S(-m_H^2,-m_H^2)|^2 
   = 1 + 0.0845 - 0.0015 + \dots \,,
\end{equation}
which indeed exhibits a vastly better behavior. 

In the expressions above, the running coupling is evaluated at time-like momentum transfer $-\mu^2-i\epsilon$. The function $\alpha_s(\mu^2)$ in perturbation theory is analytic in the complex $\mu^2$ plane with a (physical) cut on the negative real axis and a (unphysical) Landau pole at $\mu^2=\Lambda_{\overline{\rm MS}}^2$. Since we are interested in very large $|\mu^2|$ values, the Landau pole is not of concern to our discussion. The definition 
\begin{equation}\label{bdef}
   \beta(\alpha_s) = 2\,\frac{d\alpha_s(\mu^2)}{d\ln\mu^2}
   = - 2\alpha_s \sum_{n=0}^\infty \beta_n 
    \left( \frac{\alpha_s}{4\pi} \right)^n
\end{equation}
of the QCD $\beta$-function implies that
\begin{equation}
   \int_{\alpha_s(\mu^2)}^{\alpha_s(-\mu^2)}\!
   \frac{d\alpha}{\beta(\alpha)}
   = - \frac{i\pi}{2} \,,
\end{equation}
and this relation allows us to define the running coupling at time-like argument in terms of that at space-like momentum transfer. At NLO we obtain
\begin{equation}\label{asNLO}
   \frac{\alpha_s(\mu^2)}{\alpha_s(-\mu^2)}
   = 1 - ia(\mu^2) 
   + \frac{\beta_1}{\beta_0}\,\frac{\alpha_s(\mu^2)}{4\pi}\, 
   \ln\left[ 1 - ia(\mu^2) \right] + {\cal O}(\alpha_s^2) \,,
\end{equation}
where $a(\mu^2)=\beta_0\alpha_s(\mu^2)/4$. In standard applications of the renormalization group one would count this quantity as an ${\cal O}(1)$ parameter. Since numerically $a(m_H^2)\approx 0.2$, it is however also reasonable to count $a={\cal O}(\alpha_s)$.

\section{Resummation}

What is needed for the calculation of the Higgs-boson production cross section  is the Wilson coefficient at positive, not negative $\mu^2$, see (\ref{Hdef}). We will use the solution to the renormalization-group equation (\ref{CSevol}) to relate this coefficient to the one in (\ref{Ctimelike}). In that way the large corrections arising in the time-like region are resummed to all orders in perturbation theory. We write the solution in the form
\begin{equation}\label{Hresummed}
   H(m_H^2,\mu^2) = U(m_H^2,\mu^2)\,|C_S(-m_H^2,-\mu^2)|^2 \,,
\end{equation}
where \cite{Neubert:2004dd}
\begin{eqnarray}\label{Ustraight}
   \ln U(m_H^2,\mu^2)
   &=& 2\,\mbox{Re} \bigg[
    2 S(-\mu^2,\mu^2) - a_{\gamma^S}(-\mu^2,\mu^2) \nonumber\\
   &&\mbox{}- a_\Gamma(-\mu^2,\mu^2)\,\ln\frac{m_H^2}{\mu^2} 
    \bigg] \,,
\end{eqnarray}
with 
\begin{equation}
\begin{aligned}
   S(-\mu^2,\mu^2) 
   &= - \int\limits_{\alpha_s(-\mu^2)}^{\alpha_s(\mu^2)}\!
    d\alpha\,\frac{\Gamma_{\rm cusp}^A(\alpha)}{\beta(\alpha)}
    \int\limits_{\alpha_s(-\mu^2)}^\alpha
    \frac{d\alpha'}{\beta(\alpha')} \,, \\
   a_\Gamma(-\mu^2,\mu^2) 
   &= - \int\limits_{\alpha_s(-\mu^2)}^{\alpha_s(\mu^2)}\!
    d\alpha\,\frac{\Gamma_{\rm cusp}^A(\alpha)}{\beta(\alpha)} \,, 
\end{aligned}
\end{equation}
and similarly for the function $a_{\gamma^S}$. The perturbative expansions of these functions obtained at NNLO in renormalization-group improved perturbation theory can be found in \cite{Becher:2006mr}. They can be simplified using relation (\ref{asNLO}). To leading order we find
\begin{eqnarray}\label{beauty}
   \ln U(m_H^2,\mu^2)
   &=& \frac{\Gamma_0^A}{2\beta_0^2}\,\bigg\{
    \frac{4\pi}{\alpha_s(m_H^2)}\! 
    \left[ 2a\arctan(a) - \ln(1+a^2) \right] \nonumber\\
   &&\hspace{-0.8cm}\mbox{}+ \left( \frac{\Gamma_1^A}{\Gamma_0^A} 
    - \frac{\beta_1}{\beta_0} - \frac{\gamma_0^S\beta_0}{\Gamma_0^A}
    \right) \ln(1+a^2) \\
   &&\hspace{-0.8cm}\mbox{}+ \frac{\beta_1}{4\beta_0}
    \left[ 4\arctan^2(a) - \ln^2(1+a^2) \right] 
    + {\cal O}(\alpha_s) \bigg\} \,, \nonumber
\end{eqnarray}
where $a\equiv a(m_H^2)$. Note that the result is $\mu$-independent at this order. The relevant anomalous-dimension coefficients are $\Gamma_0^A=4C_A$, $\gamma_0^S=0$, and
\begin{equation}
   \frac{\Gamma_1^A}{\Gamma_0^A} 
   = \left( \frac{67}{9} - \frac{\pi^2}{3} \right) C_A
   - \frac{20}{9}\,T_F n_f \,,
\end{equation}
where $C_A=N_c$, $T_F=1/2$, and $n_f=5$ is the number of light quark flavors. The coefficients of the $\beta$-function follow from (\ref{bdef}).

The expression for the evolution function simplifies considerably if we treat $a(m_H^2)\approx 0.2$ as a parameter of order $\alpha_s$. Inserting the values of the one-loop anomalous dimensions from above, we then find
\begin{equation}\label{Ureexpanded}
   \ln U(m_H^2,\mu^2) 
   = \frac{C_A\pi\alpha_s(m_H^2)}{2}\! \left[
   1 + \frac{\Gamma_1^A}{\Gamma_0^A}\,\frac{\alpha_s(m_H^2)}{4\pi}
   + {\cal O}(\alpha_s^2) \right] .
\end{equation}
This result makes explicit that the ``$\pi^2$-enhanced'' corrections are terms of the form $(C_A\pi\alpha_s)^n$ in perturbation theory and exponentiate at leading order. The simplest way to implement our resummation in existing codes for Higgs-boson production would be to multiply the fixed-order result with $\exp[C_A\pi\alpha_s(m_H^2)/2]$ and subtract the expanded form of this factor from the perturbative series. This treatment is sufficient for practical purposes. 

Numerically, setting $\mu=m_H=120$\,GeV we obtain $\ln U=\{0.563,0.565,0.565\}$ at LO, NLO, and NNLO from the exact expression for the evolution function derived from (\ref{Ustraight}), indicating that the leading-order terms give by far the dominant effect after renormalization-group improvement. The analytical expressions (\ref{beauty}) and (\ref{Ureexpanded}) provide accurate approximations to the exact results. The first equation gives $\ln U=0.562$, while the second one yields $\ln U=0.567$. The close agreement of these two numbers shows that the running of coupling constant between $\mu^2$ and $-\mu^2$ is a minor effect compared with the evolution driven by the anomalous dimension of the effective two-gluon operator in (\ref{CSdef}).

\begin{figure}
\includegraphics[width=0.45\columnwidth]{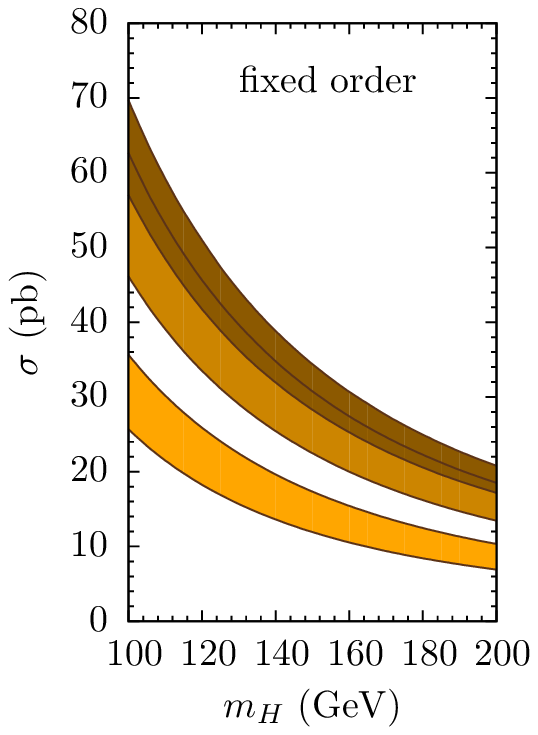} \quad
\includegraphics[width=0.45\columnwidth]{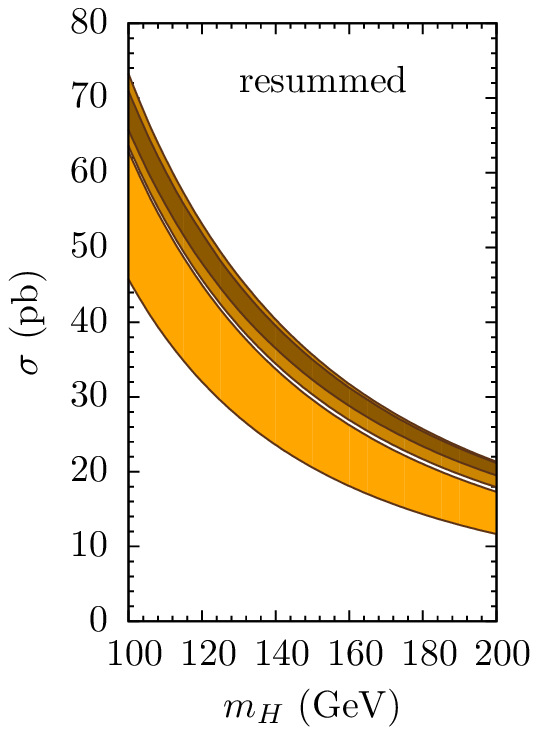}
\caption{\label{fig:sigma}
LO (light), NLO (medium), and NNLO (dark) predictions for the Higgs-production cross section at the LHC in fixed-order perturbation theory (left) and after resummation of the $\pi^2$-enhanced terms (right).}
\vspace{-0.4cm}
\end{figure}

We are now in a position to discuss our improved results for the hard function in the formula for the Higgs-boson production cross section. Setting $\mu=m_H=120$\,GeV, we obtain
\begin{equation}
   H(m_H^2,m_H^2) = \{ 1.756_{\rm \,(LO)}, 1.907_{\rm \,(NLO)},
   1.906_{\rm \,(NNLO)} \} \,.
\end{equation}
This should be compared with the poorly converging series $H=\{1,1.623,1.844\}$ obtained using fixed-order perturbation theory. Figure~\ref{fig:sigma} illustrates the impact of the resummation of the $\pi^2$-enhanced terms on the cross-section predictions for Higgs-boson production at the LHC. The bands in each plot show results obtained at LO, NLO, and NNLO using MRST2004 parton distributions \cite{Martin:2004ir}. Their width reflects the scale variation obtained by varying the factorization and renormalization scales between $m_H/2$ and $2m_H$ (setting $\mu_r=\mu_f$). The convergence of the expansion and the residual scale dependence at NLO and NNLO are much improved by the resummation. The new LO and NLO bands almost coincide with the NLO and NNLO bands in fixed-order perturbation theory, and the new NNLO band is now fully contained inside the NLO band.

\section{Drell-Yan production}

The cross section for the Drell-Yan process receives the same type of $\pi^2$-enhanced corrections as the Higgs-boson production cross section, however in this case no anomalously large $K$-factors arise at NLO and NNLO. Let us briefly discuss why this is the case. 

The vector-current matching coefficient $C_V$ appearing in the Drell-Yan case is defined in analogy with $C_S$ in (\ref{CSdef}), but with the two-gluon operator replaced by the electromagnetic current $\bar q\gamma^\mu q$ \cite{Becher:2006nr,Idilbi:2006dg,Becher:2007ty}. It obeys an evolution equation of the same structure as (\ref{CSevol}), in which the cusp anomalous dimension in the adjoint representation is replaced by that in the fundamental representation, and in which an anomalous dimension $\gamma^V$ replaces $\gamma^S$. The cusp anomalous dimensions are simply related by $\Gamma_{\rm cusp}^F/\Gamma_{\rm cusp}^A=C_F/C_A$ \cite{Vogt:2004mw}. The one-loop coefficient of $\gamma^V$ is $\gamma_0^V=-6C_F$. The resummation of $\pi^2$-enhanced terms is accomplished as in (\ref{Hresummed}), where in the explicit forms of the evolution function the appropriate expansion coefficients must be used.

The one-loop matching contribution to the Wilson coefficient $C_V$ reads
\begin{equation}\label{CVres}
   c_1^V(L) 
   = - \frac{\Gamma_0^F}{4}\,L^2 - \frac{\gamma_0^V}{2}\,L 
   + C_F \left( \frac{\pi^2}{6} - 8 \right) ,
\end{equation}
which has the same structure as (\ref{c1general}). However, the leading $\pi^2$-enhanced terms in Drell-Yan production are smaller than those in Higgs-boson production by a factor 4/9, and the constant $-8$ in the matching condition (\ref{CVres}), which is absent in the Higgs case, has the effect of partially compensating the large, positive $\pi^2$ terms resulting from the analytic continuation $L\to\ln(q^2/\mu^2)-i\pi$. This can also be seen by looking at some numerical values. For the squared coefficients at time-like momentum transfer, we obtain for $q=120$\,GeV the perturbative expansions
\begin{equation}
\begin{aligned}
   |C_V(-q^2,q^2)|^2 
   &= 1 + 0.0845 + 0.0292 + \dots \,, \\
   |C_V(-q^2,-q^2)|^2 
   &= 1 - 0.1451 - 0.0012 + \dots \,,
\end{aligned}
\end{equation} 
which should be compared with (\ref{Hres}) and (\ref{Cgood}). While the convergence is better for $\mu^2=-q^2$, the large one-loop correction seen in the Higgs case is absent for the reasons mentioned above.

It was shown in \cite{Magnea:1990zb} that the bulk of the perturbative corrections to the Drell-Yan cross section are related to $\pi^2$-enhanced terms in the ratio of the time-like to space-like Sudakov form factors. In this paper a resummation formula was derived for the ratio of these form factors, which shares some similarities with our results (\ref{beauty}) and (\ref{Ureexpanded}). To the best of our knowledge, a corresponding analysis has not been performed for the Higgs-boson production cross section. We expect that the approach based on effective field theory presented here can be adapted straightforwardly to other processes.

\vspace{-0.5cm}
\section{Conclusions}

We have shown that the large $K$-factor for Higgs-boson production at hadron colliders results from a simple kinematical effect: the analytic continuation of the gluon form factor to time-like momentum transfer. This leads to large perturbative corrections of order $(C_A\pi\alpha_s)^n$, which can be resummed to all orders by solving a renormalization-group equation. Our approach employs methods from effective field theory and properties of the QCD running coupling in the complex momentum plane. 

After the resummation of the $\pi^2$-enhanced terms, the $K$-factor for Higgs-boson production at the LHC is reduced to a rather modest value of about 1.3 at both NLO and NNLO. A detailed analysis of the phenomenological consequences of our observation, combined with state-of-the-art results for NNLO corrections and soft-gluon resummation, will be presented in a forthcoming paper \cite{inprep}. Extensions of our approach to other hard-scattering processes with time-like momentum transfer, such as event shapes or $t\bar t$ production at hadron colliders, will be discussed elsewhere. 

\vspace{0.1cm}
{\em Acknowledgments:\/}
We are grateful to Babis Anastasiou, Martin Beneke, Frank Petriello and Giulia Zanderighi for useful discussions. The research of T.B.\ was supported by the U.S.\ Department of Energy under Grant DE-AC02-76CH03000. Fermilab is operated by the Fermi Research Alliance under contract with the Department of Energy.


\begin{thebibliography}{99}

\bibitem{Dawson:1990zj}
  S.~Dawson,
  Nucl.\ Phys.\  B {\bf 359}, 283 (1991).

\bibitem{Djouadi:1991tka}
  A.~Djouadi, M.~Spira and P.~M.~Zerwas,
  Phys.\ Lett.\  B {\bf 264}, 440 (1991).

\bibitem{Harlander:2002wh}
  R.~V.~Harlander and W.~B.~Kilgore,
  Phys.\ Rev.\ Lett.\  {\bf 88}, 201801 (2002)
  [arXiv:hep-ph/0201206].

\bibitem{Anastasiou:2002yz}
  C.~Anastasiou and K.~Melnikov,
  Nucl.\ Phys.\  B {\bf 646}, 220 (2002)
  [arXiv:hep-ph/0207004].

\bibitem{Ravindran:2003um}
  V.~Ravindran, J.~Smith and W.~L.~van Neerven,
  Nucl.\ Phys.\  B {\bf 665}, 325 (2003)
  [arXiv:hep-ph/0302135].

\bibitem{Catani:2003zt}
  S.~Catani, D.~de Florian, M.~Grazzini and P.~Nason,
  JHEP {\bf 0307}, 028 (2003)
  [arXiv:hep-ph/0306211].

\bibitem{Inami:1982xt}
  T.~Inami, T.~Kubota and Y.~Okada,
  Z.\ Phys.\  C {\bf 18}, 69 (1983).

\bibitem{Chetyrkin:1997iv}
  K.~G.~Chetyrkin, B.~A.~Kniehl and M.~Steinhauser,
  Phys.\ Rev.\ Lett.\  {\bf 79}, 353 (1997)
  [arXiv:hep-ph/9705240].

\bibitem{Becher:2006nr}
  T.~Becher and M.~Neubert,
  Phys.\ Rev.\ Lett.\  {\bf 97}, 082001 (2006)
  [arXiv:hep-ph/0605050].

\bibitem{Idilbi:2006dg}
  A.~Idilbi, X.~d.~Ji and F.~Yuan,
  Nucl.\ Phys.\  B {\bf 753}, 42 (2006)
  [arXiv:hep-ph/0605068].

\bibitem{Becher:2007ty}
  T.~Becher, M.~Neubert and G.~Xu,
  JHEP {\bf 0807}, 030 (2008)
  [arXiv:0710.0680 [hep-ph]].

\bibitem{Bauer:2001yt}
  C.~W.~Bauer, D.~Pirjol and I.~W.~Stewart,
  Phys.\ Rev.\  D {\bf 65}, 054022 (2002)
  [arXiv:hep-ph/0109045].

\bibitem{Bauer:2002nz}
  C.~W.~Bauer, S.~Fleming, D.~Pirjol, I.~Z.~Rothstein and I.~W.~Stewart,
  Phys.\ Rev.\  D {\bf 66}, 014017 (2002)
  [arXiv:hep-ph/0202088].

\bibitem{Hill:2002vw}
  R.~J.~Hill and M.~Neubert,
  Nucl.\ Phys.\  B {\bf 657}, 229 (2003)
  [arXiv:hep-ph/0211018].

\bibitem{Harlander:2000mg}
  R.~V.~Harlander,
  Phys.\ Lett.\  B {\bf 492}, 74 (2000)
  [arXiv:hep-ph/0007289].

\bibitem{inprep}
V.~Ahrens, T.~Becher, M. Neubert and L.~L.~Yang, in preparation.

\bibitem{Vogt:2004mw}
  A.~Vogt, S.~Moch and J.~A.~M.~Vermaseren,
  Nucl.\ Phys.\  B {\bf 691}, 129 (2004)
  [arXiv:hep-ph/0404111].

\bibitem{Moch:2005tm}
  S.~Moch, J.~A.~M.~Vermaseren and A.~Vogt,
  Phys.\ Lett.\  B {\bf 625}, 245 (2005)
  [arXiv:hep-ph/0508055].

\bibitem{Neubert:2004dd}
  M.~Neubert,
  Eur.\ Phys.\ J.\  C {\bf 40}, 165 (2005)
  [arXiv:hep-ph/0408179].

\bibitem{Becher:2006mr}
  T.~Becher, M.~Neubert and B.~D.~Pecjak,
  JHEP {\bf 0701}, 076 (2007)
  [arXiv:hep-ph/0607228].

\bibitem{Martin:2004ir}
  A.~D.~Martin, R.~G.~Roberts, W.~J.~Stirling and R.~S.~Thorne,
  Phys.\ Lett.\  B {\bf 604}, 61 (2004)
  [arXiv:hep-ph/0410230].

\bibitem{Magnea:1990zb}
  L.~Magnea and G.~Sterman,
  Phys.\ Rev.\  D {\bf 42}, 4222 (1990).

\end{thebibliography}
\end{document}